\documentclass[ prd,twocolumn]{revtex4}
\usepackage{xcolor}
\usepackage[utf8]{inputenc}
\usepackage{graphicx}
\usepackage{empheq}
\usepackage{braket}
\usepackage{float}
\usepackage{amssymb}
\usepackage{amsmath}
\usepackage{mhchem}
\usepackage{siunitx}
\usepackage[titletoc]{appendix}

\usepackage{hyperref}
\hypersetup{colorlinks=true, linkcolor=red, citecolor=red, urlcolor=purple}

\newcommand\purple[1]{{\color{purple}#1}}

\begin{document}

\title{\purple{Vortex state in a superconducting mesoscopic irregular octagon}}
\author{C. A. Aguirre$^{1,2,}$\footnote[2]{\href{cristian@fisica.ufmt.br}{cristian@fisica.ufmt.br}}\href{https://orcid.org/0000-0001-8064-6351}{\includegraphics[scale=0.05]{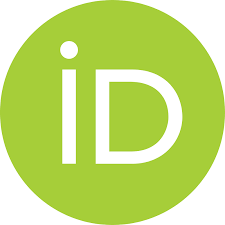}}}
\author{Julián Faúndez$^{2}$\href{https://orcid.org/0000-0002-6909-0417}{\includegraphics[scale=0.05]{Figures/orcid.png}}}
\author{J. Barba-Ortega$^{3,4}$\href{https://orcid.org/0000-0003-3415-1811}{\includegraphics[scale=0.05]{Figures/orcid.png}}}
\affiliation{$^{1}$Departamento de Física, Universidade Federal de Mato-Grosso, Cuiabá, Brasil}
\affiliation{$^{2}$Condensed Matter Physics Group, Instituto de Fisica, Universidade Federal do Rio Grande do Sul, $91501-970$ Porto Alegre, RS, Brazil.}
\affiliation{$^{3}$Departamento de Física, Universidad Nacional de Colombia, Bogotá, Colombia}
\affiliation{$^{4}$Foundation of Researchers in Science and Technology of Materials, Bucaramanga, Colombia}
\date{\today}
\begin{abstract} 
Our study sample is a superconducting bi-dimensional octagon with different boundary conditions immersed in a magnetic external field $H$. The boundary conditions are simulated by considering different values of the deGennes extrapolation length $b$ on different surfaces of the sample. Our investigation was carried out by solving the two-band time dependent Ginzburg–Landau equations (TB-TDGL). We analyzed the superconducting electron density and the magnetization curves as functions of $H$ and temperature $T$ in zero-field cooling and in zero-field-cooling processes for different values of $b$ and size of the sample. We found a strong dependence of the critical fields on $b$ and size of the sample.
\end{abstract}
\maketitle
\section{Introduction}
The technological applications for high critical temperature superconducting materials in different areas of the sciences have generated great interest in the last decades \cite{1,2,3,4,5}. This field has developed since its discovery in 1908 by H. Onnes, G. Holts and J. Flint \cite{6}, encompassing different branches that they contemplate, mesoscopic superconductivity \cite{7,8}, topological \cite{9,10,11}, multi-band \cite{12,13}, frustrated superconductivity \cite{13a} and its applications in electronic devices reaffirm its importance in the advancement of new technologies based on vortex control \cite{14,15}, Magnons\cite{15a}  or Hopfios\cite{15b}, not without first mentioning the discovery of superconductivity in ferromagnetic systems \cite{15c}. In general, until now, there are three specific theoretical models for to study the superconducting state. The first is based on principles of many-body quantum mechanics in conjunction with some experimental data and is called Ab-initio \cite{15d,15e}, the method is mainly used, perturbative-DFT \cite{15f,15g}, the main deficiency of this method is the null capacity for the study of vortices and other pseudo-particles. The second method, based on extensions of the Bardeen-Cooper-Schrieffer (BCS) theory, has the ability to study superconductivity on a mesoscopic scale, giving excellent results compared to experimental ones \cite{15h,15i}, but it neglects the geometry of the sample itself, besides that its solution is based mainly on the diagonalization of the Bogoliubov de Gennes Hamiltonian, a main problem is that it presents long computation times \cite{15j}. Now, the third form of study is based on the extension of the second order transition theory, proposed by L. Landau to superconducting systems, accounting for a parameter of order $\psi$ or pseudo-wave function that describes the superconducting state in the mesoscopic regime ($d \approx \xi(0)$) \cite{15k,15l}. The latter is widely used for the study of the dynamics of vortices in different geometries, which are mainly regular domains (circles, squares/rectangles or triangles) \cite{15m,15n,15o}, due to the high convergence times and the difficulty of applying boundary conditions in domains with geometries different from those mentioned above, little or no other types of geometries are discussed.
That is why, in the present work, we will study the vortex state and magnetization, for a superconducting bi-dimensional octagon. Then, we will study the effect that different boundary conditions have on these variables. Finally, we will extend the model to address a two-condensate system, mediated by a Josephson-type coupling between them and we will study the first stable vortex configuration, for both condensates. The study of this system will be based on the solution of the two-condensates Ginzburg-Landau time-dependent equations (TB-TDGL).\\\\
This article is organized as follows: the theoretical formalism is presented in section \ref{Section1}. In section \ref{Section2}, \ref{Section2a} and \ref{Section2b}, we present the main results for the studied system, we show the vortex states and magnetization. Finally, in section \ref{Section3} we detail the main results.
\section{Theoretical Formalism}\label{Section1}
\begin{figure}[H]
    \centering
    \includegraphics[scale=0.4]{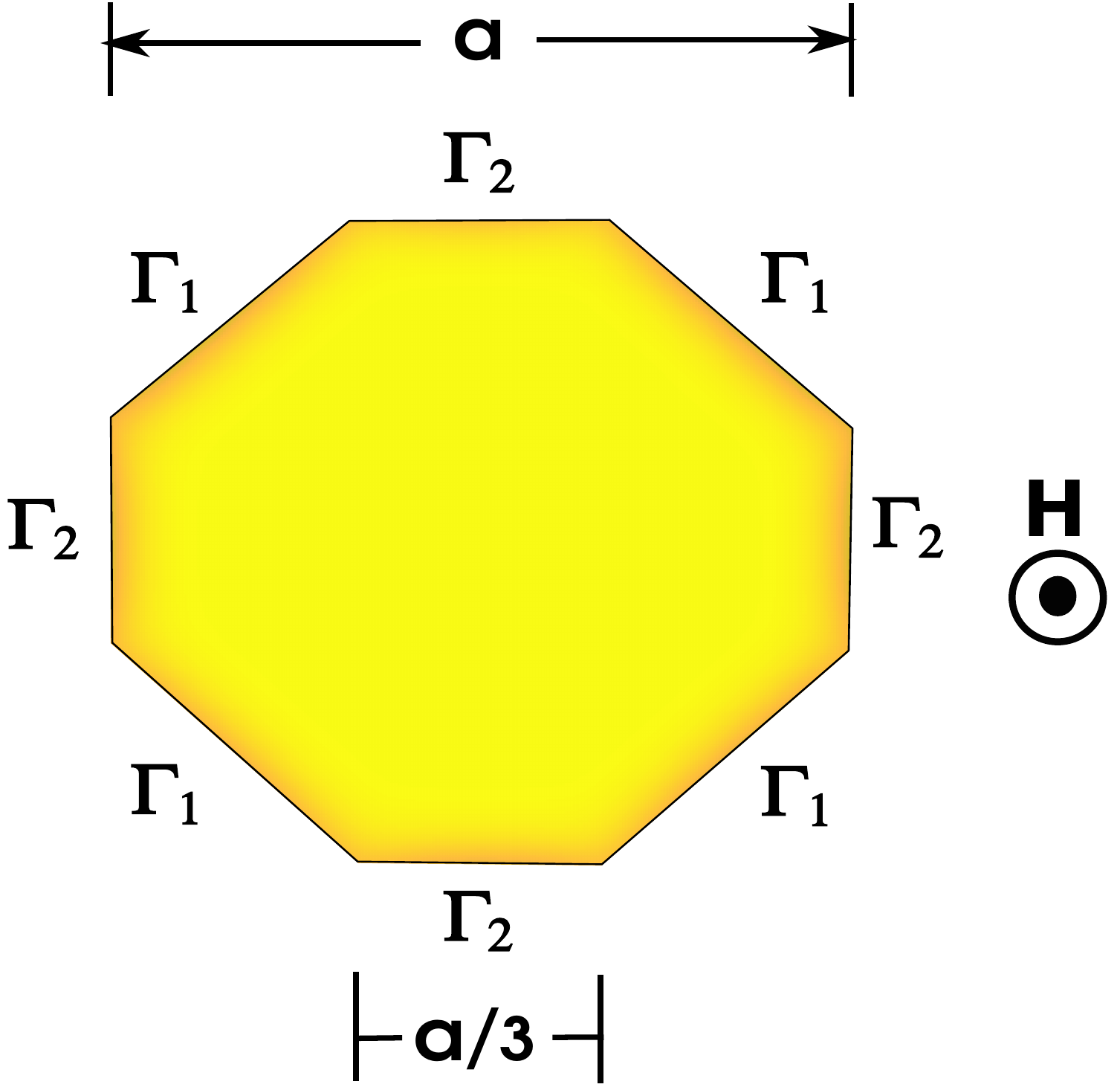}
   \caption{Layout of the studied sample immersed in a perpendicular magnetic field $\mathbf{H}$, the octagon area is $A=7a^{2}\xi^{2}/9$.}
    \label{Layout}
\end{figure}

In the present work, we will study the superconducting state for a mesoscopic octagon (see Fig. \ref{Layout}). We will consider several sizes of the sample $a/\xi=10,20,30$. ($a$ is the side of the square that submit the octagon (see Fig.  \ref{Layout}). Additionally, the borders are in contact  with different materials. 
With this, we will consider the Gibbs functional for a two-condensate with a Josephson coupling,  where $\psi_{i}$ is the superconducting order parameter complex pseudo-function $\psi_{i}=|\psi_{i}|e^{i\theta_{i}}$, ($\theta_{i}$ its phase) for the $i=1,2$ condensate. \cite{16,17}, $\mathbf{A}$ is the magnetic potential:
\begin{eqnarray}
\mathcal{G} = \int dV ( \sum_{i}^{2} \mathcal{F}(\psi_{i},\textbf{A})
+\frac{1}{2 \mu_0} |\nabla \times \mathbf{A}|^2+\Xi(\psi_{i}))\label{Gibbs1}
\end{eqnarray}
with:
{\small
\begin{equation}
\mathcal{F}(\psi_{i},\textbf{A})= \alpha_i |\psi_i|^2 +\frac{\beta_i}{2} |\psi_i|^4+\frac{\zeta_{i}}{2 m_i} |(i \hbar \nabla+2 e \mathbf{A}) \psi_i |^2
\label{Function}
\end{equation}}
and
\begin{eqnarray}
\Xi(\psi_{1},\psi_{2})=\gamma (\psi_1^*\psi_2+\psi_2^*\psi_1).
\label{Jops}
\end{eqnarray}
We will consider, $\alpha_i=\alpha_{i0}(1-T/T_{ci})$ and $\beta_i$ are two phenomenological parameters, $i=1,2$ in the equations \ref{Gibbs1} and \ref{Function}. We used the Josephson coupling showed in the equation \ref{Jops}. We express the temperature $T$ in units of the critical temperature $T_{c1}$, length in units of the coherence length $\xi_{10}=\hbar/\sqrt{-2 m_1 \alpha_{10}}$, the order parameters in units of $\psi_{i0}=\sqrt{-\alpha_{i0}/\beta_i}$, time in units of the Ginzburg-Landau characteristic time $t_{GL}=\pi\hbar /8k_{B}T_{c1}$, and the vector potential $\mathbf{A}$ is scaled by $H_{c2}\xi_{10}$, where $H_{c2}$ is the bulk upper critical field. The two-condensate superconductor dynamical equations in dimensionless units is given by \cite{18,19}:
\begin{align}
\frac{\partial \psi_{1}}{\partial t}=(1-T-|\psi_1|^2) \psi_1-|\mathbf{D}|^2 \psi_1 +\gamma|\psi_{2}|e^{i\theta_{2}}
\label{GL2B1}
\end{align}
\begin{eqnarray}
\frac{\partial \psi_{2}}{\partial t}=(1-\frac{T}{T_{r2}}-|\psi_2|^2)\psi_2- 
\frac{m_{r2}}{\alpha_{r2}} |\mathbf{D}|^2 \psi_2 +\gamma|\psi_{1}|e^{i\theta_{1}}.
\label{GL2B2}
\end{eqnarray}
The Eq. (\ref{GL2B1}) and Eq. (\ref{GL2B2}) are solved in $\Omega_{sc}$. The equation for the vector potential $\mathbf{A}$:
\begin{eqnarray}
\frac{\partial{\bf A}}{\partial t}= {\bf J}_s-\kappa^2\nabla\times\nabla\times {\bf A}
\label{EQ5}
\end{eqnarray}
also are solved in $\Omega_{sc}$, where:
\begin{equation}
{\bf J}_{s} = \zeta_{1}\Re \left[\psi_1\mathbf{D}\psi_1^*  \right]+ 
\zeta_{2} \Re \left[\frac{\beta_{r2}}{\alpha_{r2}}\psi_2\mathbf{D}\psi_2^*\right].
\label{EQ6}
\end{equation}
For this case, $\zeta_{1}=\zeta_{2}$, also $\mathbf{D}=i\nabla -\mathbf{A}$, and  $\gamma$ represents the strong of the Josephson coupling between the $i$ and $j$ band. The boundary conditions ${\bf n}\cdot(i\mbox{\boldmath$\nabla$}+{\bf A})\psi_{i}=i\hbar\psi/b$, the superconductor–vacuum interface is simulated by $b\rightarrow \infty$. The case in which $b>0$ and $b\approx \xi$, is used for a superconductor-metal interface. $b<0$ is considered by a superconducting-superconductor at higher critical temperature $T_c$ interface. ${\bf n}$ outer surface normal vector. Finally, for the coupling constants $m_{r2}=m_{2}/m_{1}=0.5$, $\beta_{r2}=\beta_{2}/\beta_{1}=0.7$, $\gamma=-0.01$ (healing coupling), $\zeta_{1}=\zeta_{2}=0.01$, the Ginzburg-Landau parameter $\kappa=1.0$ \cite{19}. In the single-condensate case ($\gamma=0$) in the Eq. (\ref{Jops}), we consider the limiting case, where the coupling is null and the second parameter does not exist, with which the eq.(\ref{GL2B2}) does not exist, leaving only a superconducting condensate $\psi_{1}$. For the computational mesh we use $\Delta x=\Delta y=0.1$ \cite{13,14,15,16}. In the  field cooling processes simulations for we take $T=0.1$.
\subsection{Single-condensate case $\gamma=0$, $\Gamma=1.0$. $a$ variation.}\label{Section2}
\begin{figure}[H]
    \centering
    \includegraphics[scale=0.29]{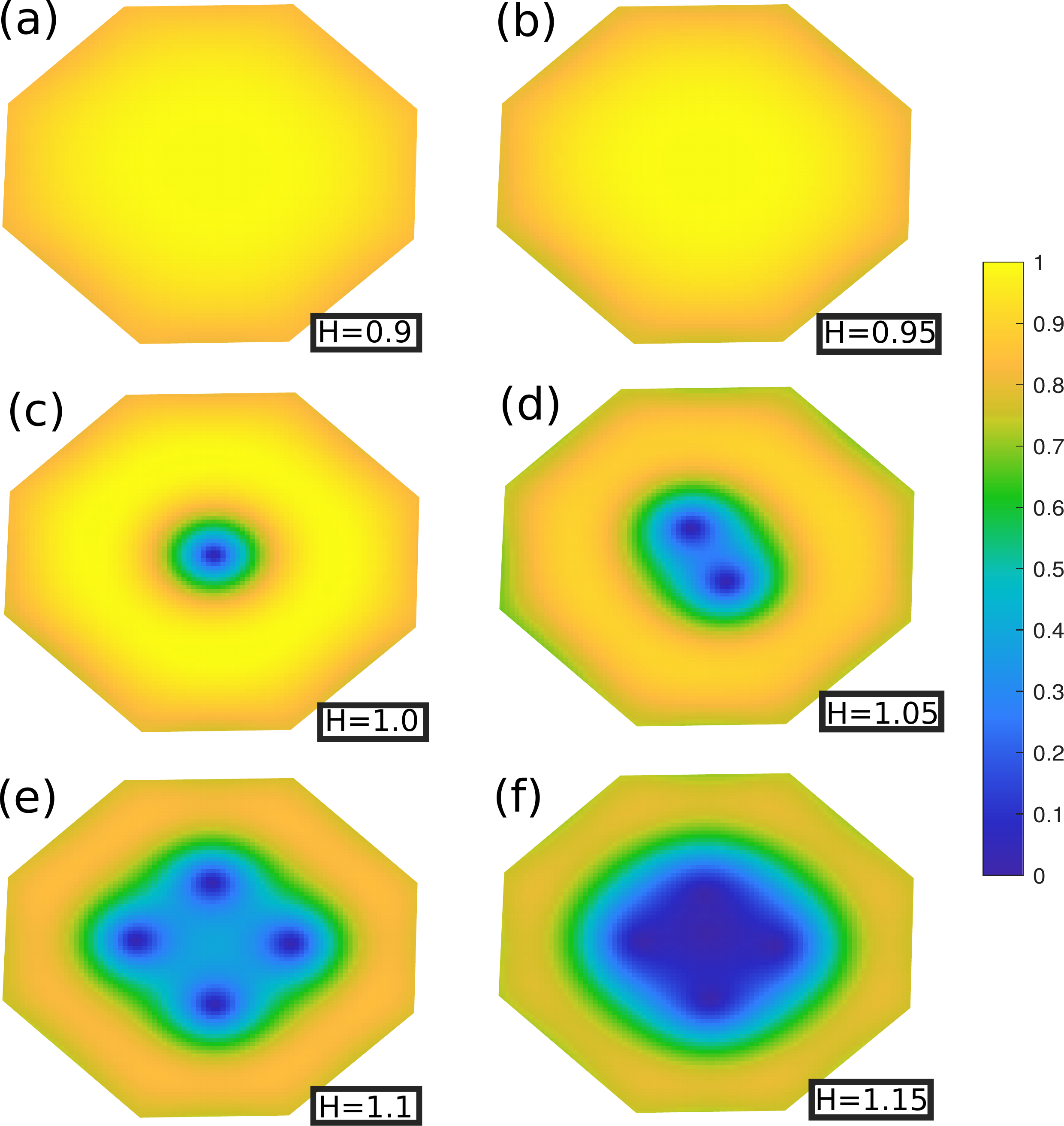}
    \caption{Cooper pairs density $|\psi|^{2}$ at the indicates $H$ for a single-condensate sample considering  $b\rightarrow \infty$, and $a=10\xi$.}
    \label{psi1band10x10}
\end{figure}

\begin{figure}[H]
    \centering
    \includegraphics[scale=0.29]{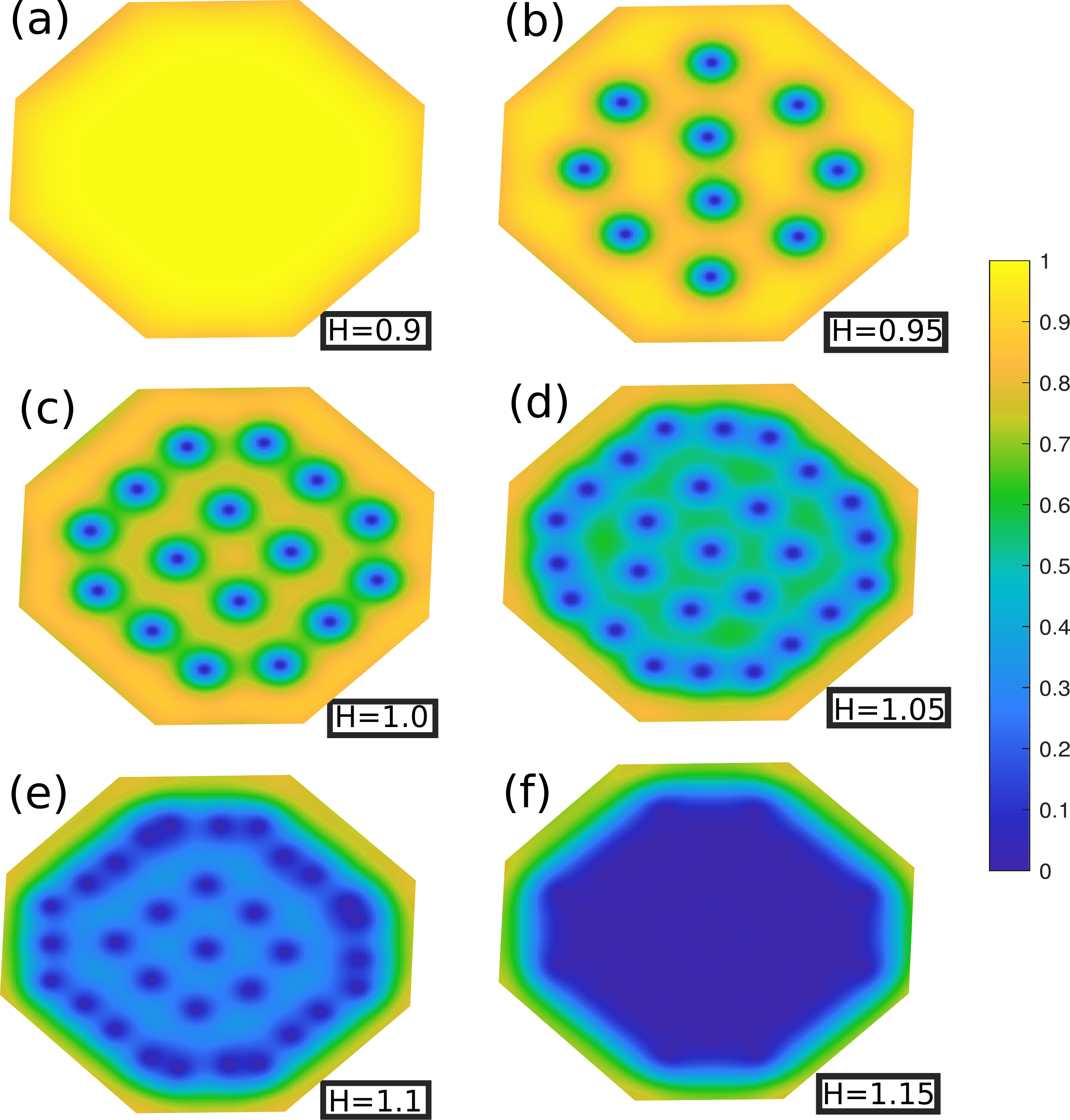}
    \caption{Square order parameter $|\psi|^{2}$ at indicates $H$ for a single-condensate, $b\rightarrow \infty$, for $a=20\xi$.}
    \label{psi1band20x20}
\end{figure}

In the Fig. (\ref{psi1band10x10}), we show the superconducting electronic density $|\psi|^{2}$ or density of Cooper pairs for a single-condensate sample, at several magnetic field  $H$ and $a=10\xi$. In the Fig. (\ref{psi1band10x10})(a) and Fig. (\ref{psi1band10x10})(b), for $H<H_1$ (where $H_1$ is the $H$ where occur the first vortex entry), $H<0.95$ we can observe that the system does not present vortices ($N=0$, $N$ is the vorticity, Meissner-Oschenfeld state). In Fig. (\ref{psi1band10x10})(c) at $H=1.0$, the breaks the Meissner state into a mixed state, where $N=1$ vortices enter into the sample. In the Fig. (\ref{psi1band10x10})(d) and Fig. (\ref{psi1band10x10})(e)), at $H=1.05$, $N=2$ and $H=1.1$, $N=4$ respectively, we observe that due to the geometry of the sample, the point of greatest symmetry is the center and it is where the first vortex that enters is located, due to this entry the energy barrier at the border decreases, favoring the possibility of the entry of new vortices. Also, due to the repulsive interaction between the vortices, the new vortices are initially mobilized by the sample (transient states), after which they are located in positions of high symmetry (stationary states), as this process is repeated over time. At $H=1.15$ the centers of the vortices coincide, leading to the formation of a giant vortex in the center of the sample  (See Fig. (\ref{psi1band10x10})(f)).
In Fig. (\ref{psi1band20x20}) we show the vortex state  at indicates $H$ considering $a=20\xi$. In  Fig. (\ref{psi1band20x20})(a) We observe the sample in the Meissner-Oschenfeld state at $H=0.9$, $N=0$. As the external field is increased $H=0.95$, the superconductor transiting to the mixed state (superconductor/normal) where $N=10$ vortices appear, (see Fig. (\ref{psi1band20x20})(b)). As the magnetic field continues to increase $H=1.0$, $N=16$ and   for $H=1.05$, $N=26$  vortices entry to the sample, (see Figs. (\ref{psi1band20x20})(c)-(d)), It is of special importance to observe the symmetry in the position of the vortices closer to the border, with which, due to the greater entry of vortices and the magnetic pressure that their repulsion generates, a giant and unique state is formed. vortex homogeneously distributed over the entire surface of the sample. It is important to note that at higher $H>1,1$, it is not possible to extract information from this graph to know the vortex number into the sample (see Fig.\ref{psi1band20x20})(e)-(f).

\begin{figure}[H]
    \centering
    \includegraphics[scale=0.3]{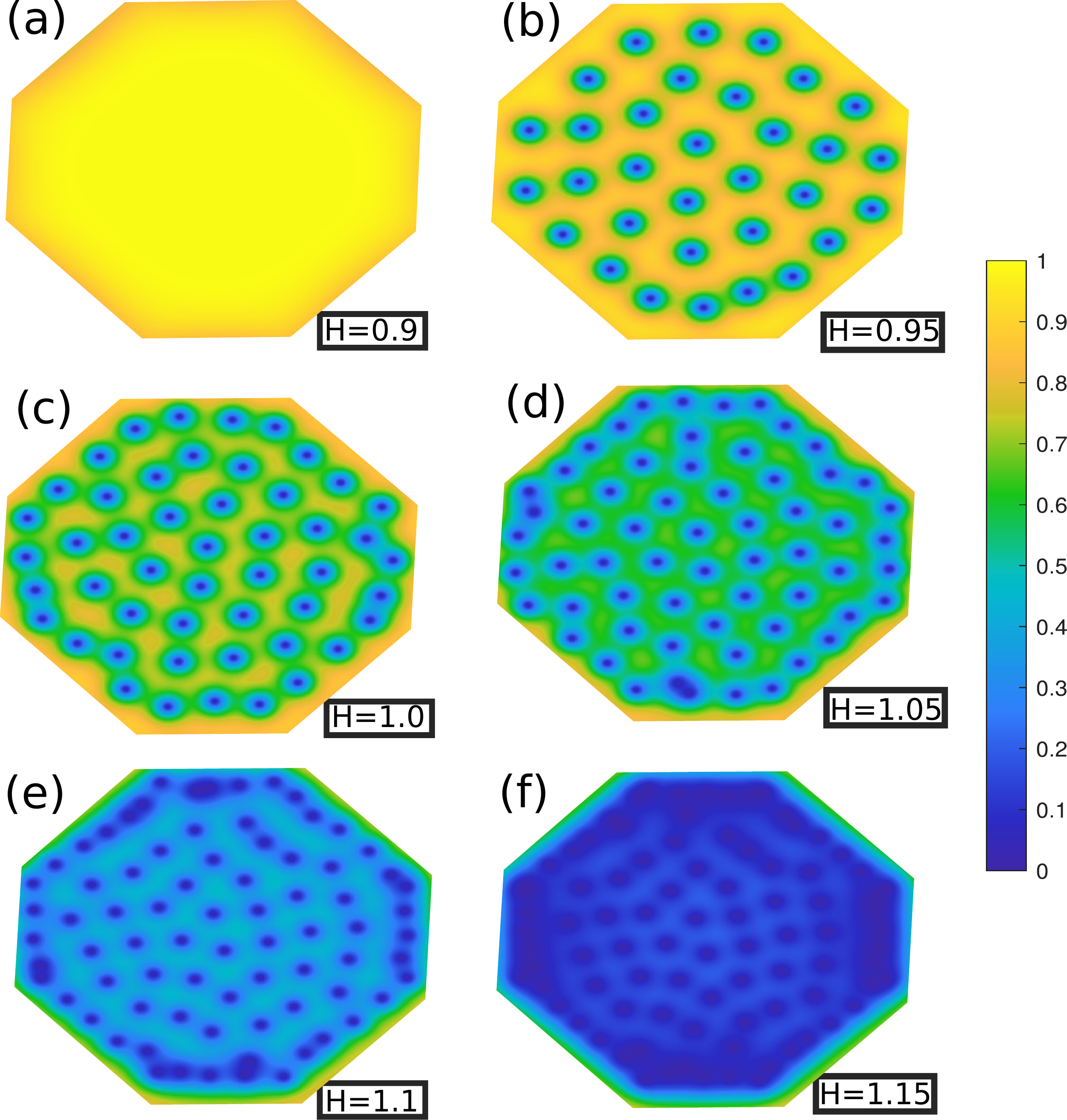}
    \caption{Square order parameter $|\psi|^{2}$ at indicates $H$ for a single-condensate, $b\rightarrow \infty$, for $a=30\xi$.}
    \label{psi1band30x30}
\end{figure}

Finally in Fig. (\ref{psi1band30x30}) we show the vortex state for $a=30\xi$. We can see in Fig. (\ref{psi1band30x30})(a), at $H>0.9$ the system is observed in the Meissner-Oschenfeld state (N=0); as $H$ increases, the conventional behavior of vortex entry begins, initially at the center of the octagon (symmetry), which increase in numerical terms, as the field continues to increase until reaching a point where the surface of the octagon has a low energy barrier. It should be noted that the formation of vortex states when $a=30\xi$ y $a=20\xi$ are more quickly caught compared to $a=10\xi$. In addition, the effects of size lead to the formation of a greater number of vortices on the surface, hoping that as the size of the sample increases, the first critical field $H_{c1}$, in which the loss of the Meissner state exists, decrease.  
\begin{figure}[H]
    \centering
    \includegraphics[scale=0.5]{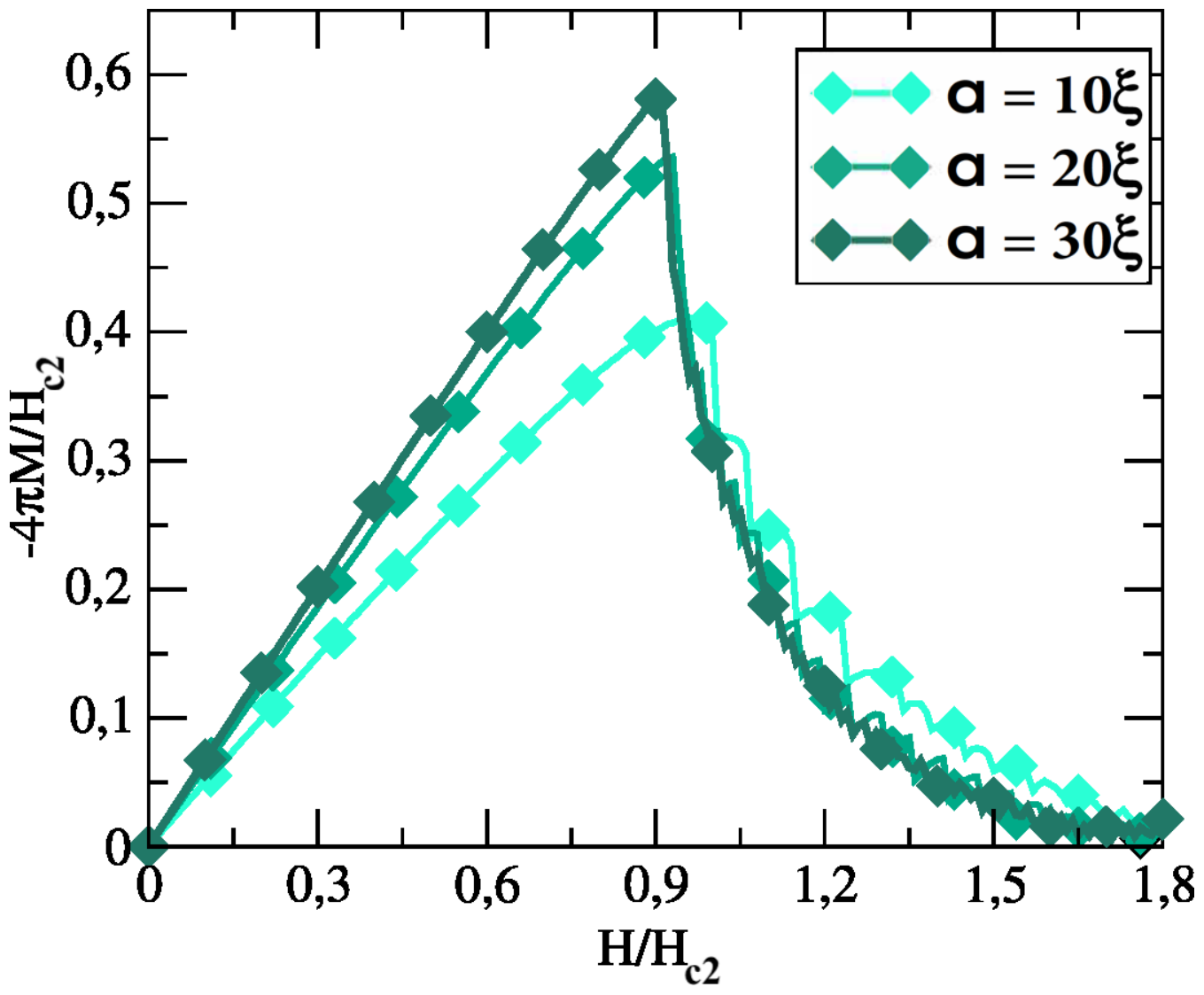}
    \caption{Magnetization $-4\pi M/H_{c2}$ as function of $H$ for $a/\xi=10,20,30$ sizes.}
    \label{M102030}
\end{figure}
In Fig. (\ref{M102030}) we present the magnetization as a function of the applied magnetic field $\mathbf{H}$ for $a/\xi=10,20,30$. It is easy to see that there are three maximums of magnetization, $-4\pi M_{max}/H_{c2}\approx 0.58$ for $a=10\xi$; $-4\pi M_{max}/H_{c2}\approx 0.52$ for $a=20\xi$; and $-4\pi M_{max}/H_{c2}\approx 0.40$ for $a=30\xi$; which present the value of maximum diamagnetism of the sample. The lower field $H_1$ depends slightly on $a$, as it was presented in the Fig. (\ref{psi1band10x10})-Fig. (\ref{psi1band30x30}), the first critical field decreases as the size of the superconducting sample increases. By another hand, the upper field  (superconducting-normal transition field) $H_2$ does not depend on $a$.
\subsection{Single-condensate case $\gamma=0$, $a=20\xi$, $\Gamma$ variation.}\label{Section2a}
\begin{figure}[H]
    \centering
    \includegraphics[scale=0.51]{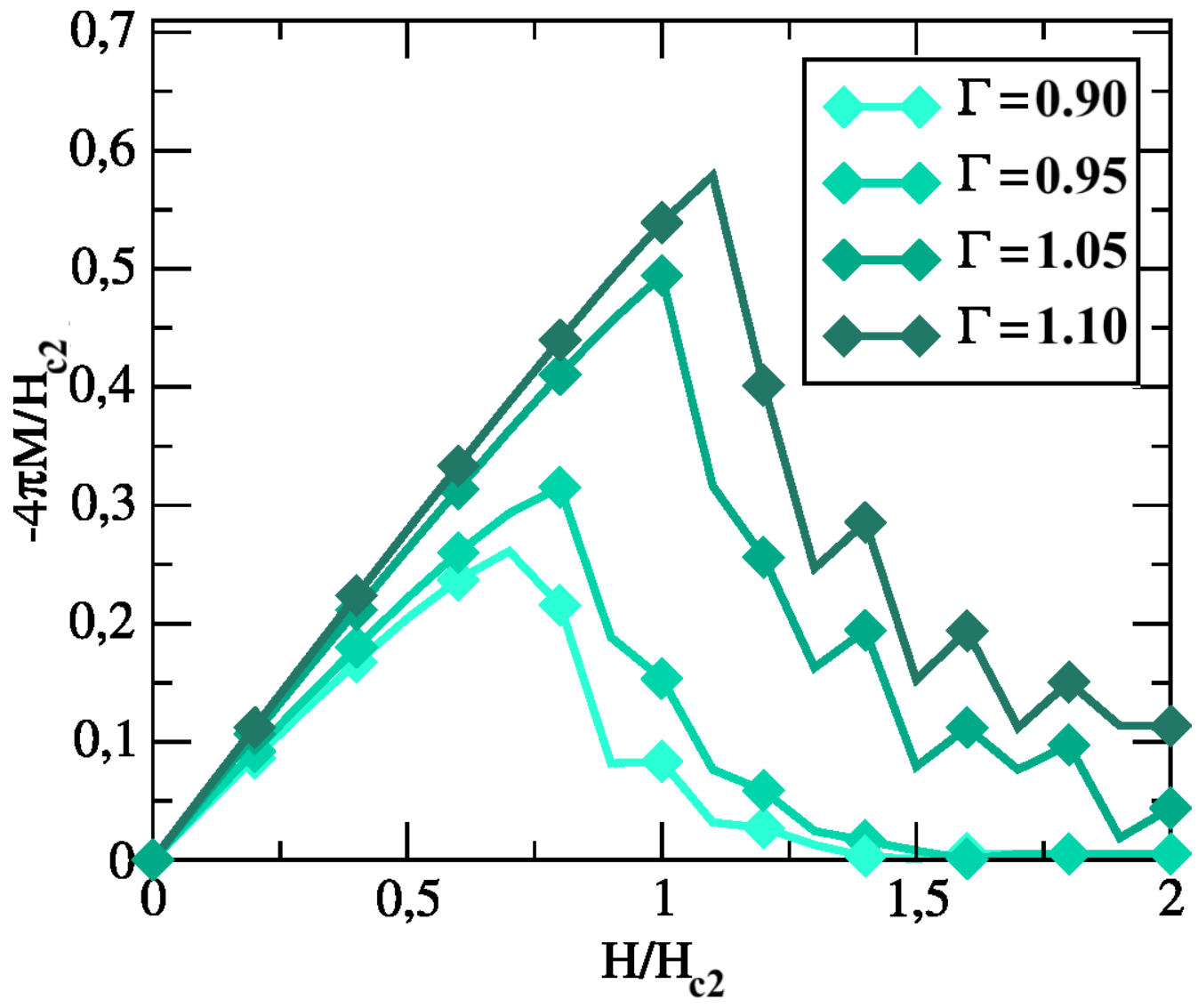}
    \caption{Magnetization $-4\pi\mathbf{M}/H_{c2}$ as function of $\mathbf{H}$ at indicates $\Gamma$ in four opposite sides and $\Gamma=1.0$ in the other four sides.}
    \label{M20x20B}
\end{figure}

\begin{figure}[H]
    \centering
    \includegraphics[scale=0.28]{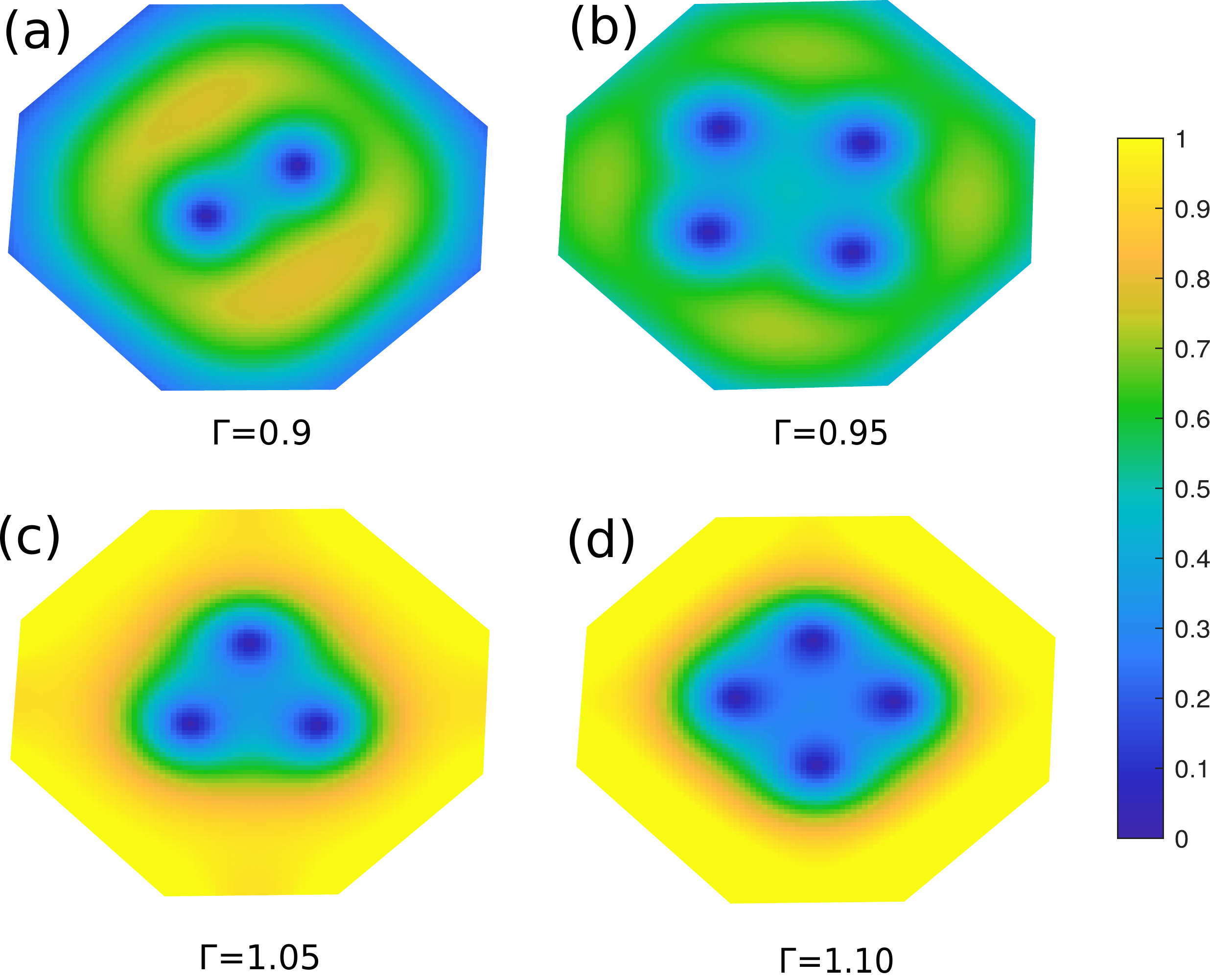}
    \caption{First stable vortex configuration for all border with the same boundary for (a) $\Gamma=0.90$ at $H_1=0.75$, (b) $\Gamma=0.95$ $H_1=0.85$, (c) $\Gamma=1.05$ at $H_1=0.95$, and d) $\Gamma=1.10$ at $H_1=0.95$ in all the sides.}
    \label{dGhomogeneo}
\end{figure}

\begin{figure}[H]
    \centering
    \includegraphics[scale=0.51]{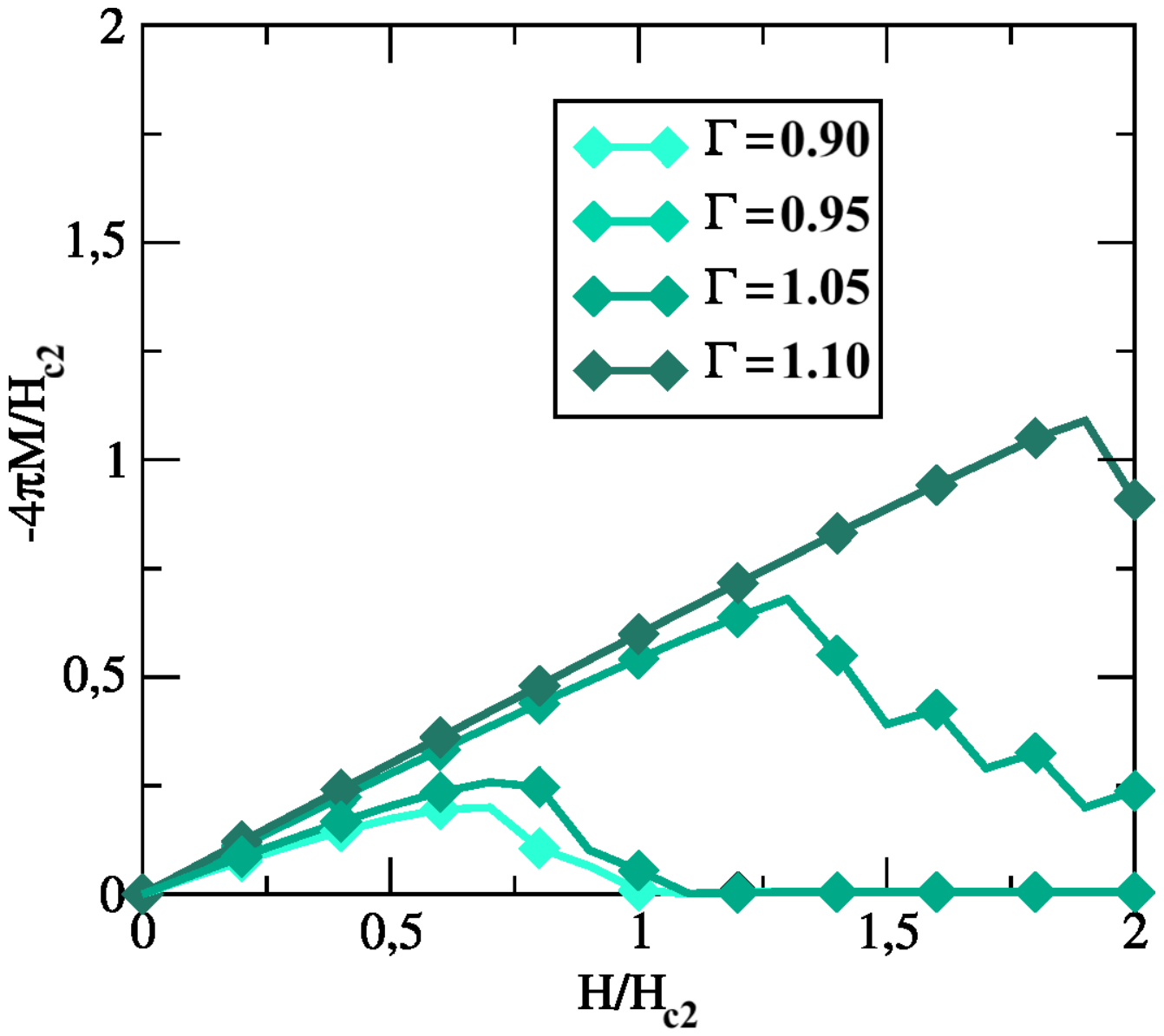}
    \caption{Magnetization $-4\pi M/H_{c2}$ as function of $H$ at indicated $\Gamma$ in the eight surfaces.}
    \label{M20x20-H}
\end{figure}

\begin{figure}[H]
    \centering
    \includegraphics[scale=0.3]{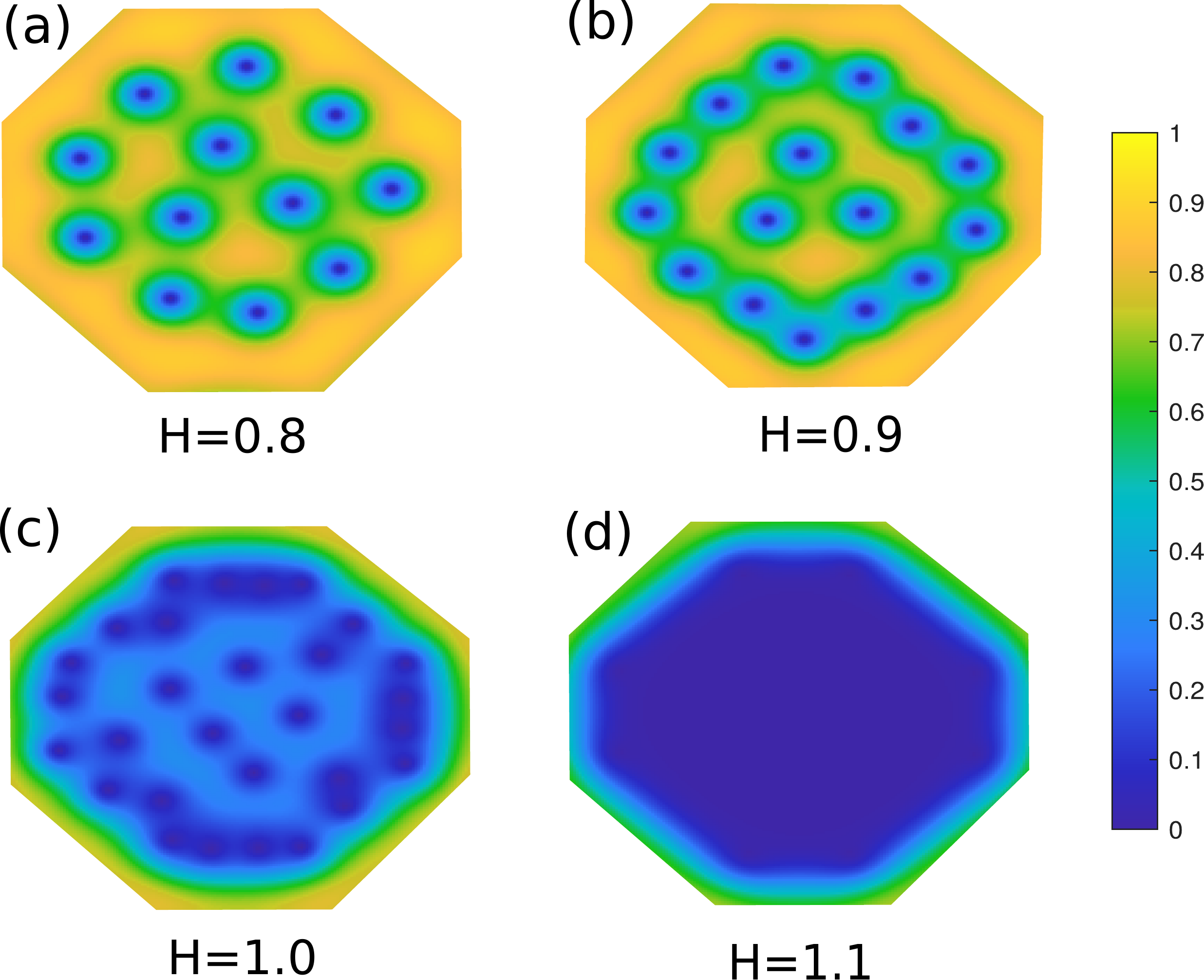}
    \caption{Square order parameter $|\psi_{i}|^{2}$ in a ZFC process at indicates $H$, and at $T=0.1$.}
    \label{ZFCVortex}
\end{figure}

\begin{figure}[H]
    \centering
    \includegraphics[scale=0.51]{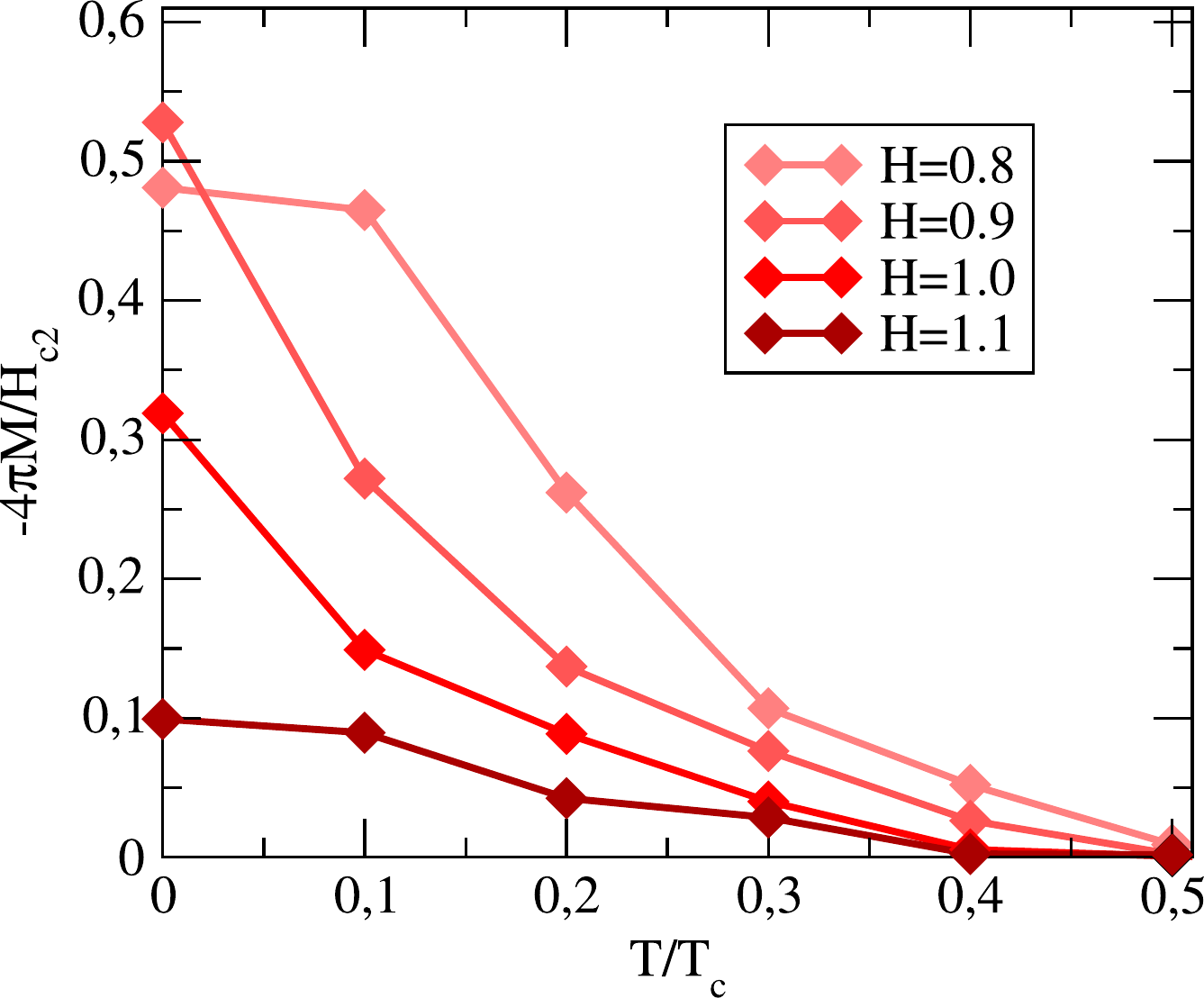}
    \caption{Magnetization $-4\pi M/H_{c2}$ for a single-component sample as function of $T/T_{c}$ for $a=20\xi$, in a ZFC process.}
    \label{MZFC}
\end{figure}

In this section we study the effect that the de Gennes parameter $b$ has on the magnetization and the vortex state in a single-condensate sample. To do this, We unify the boundary conditions upon introducing the $\Gamma= 1-\Delta/b$ parameter. Thus, $0<\Gamma<1$ simulates a superconductor/metal interface; a superconductor/superconductor at higher critical temperature is described with $\Gamma>1.0$; and $\Gamma=1.0$ simulates a superconductor/vacuum interface. We will vary the interface of four sides of the octagon (See Fig. \ref{Layout}), keeping the other interfaces invariant.  We consider $\Gamma=0.90$; $0.95$; $1.05$; $1.10$, in four opposite sides and  $\Gamma=1.0$ for the other four sides.
\begin{figure}[H]
    \centering
    \includegraphics[scale=0.29]{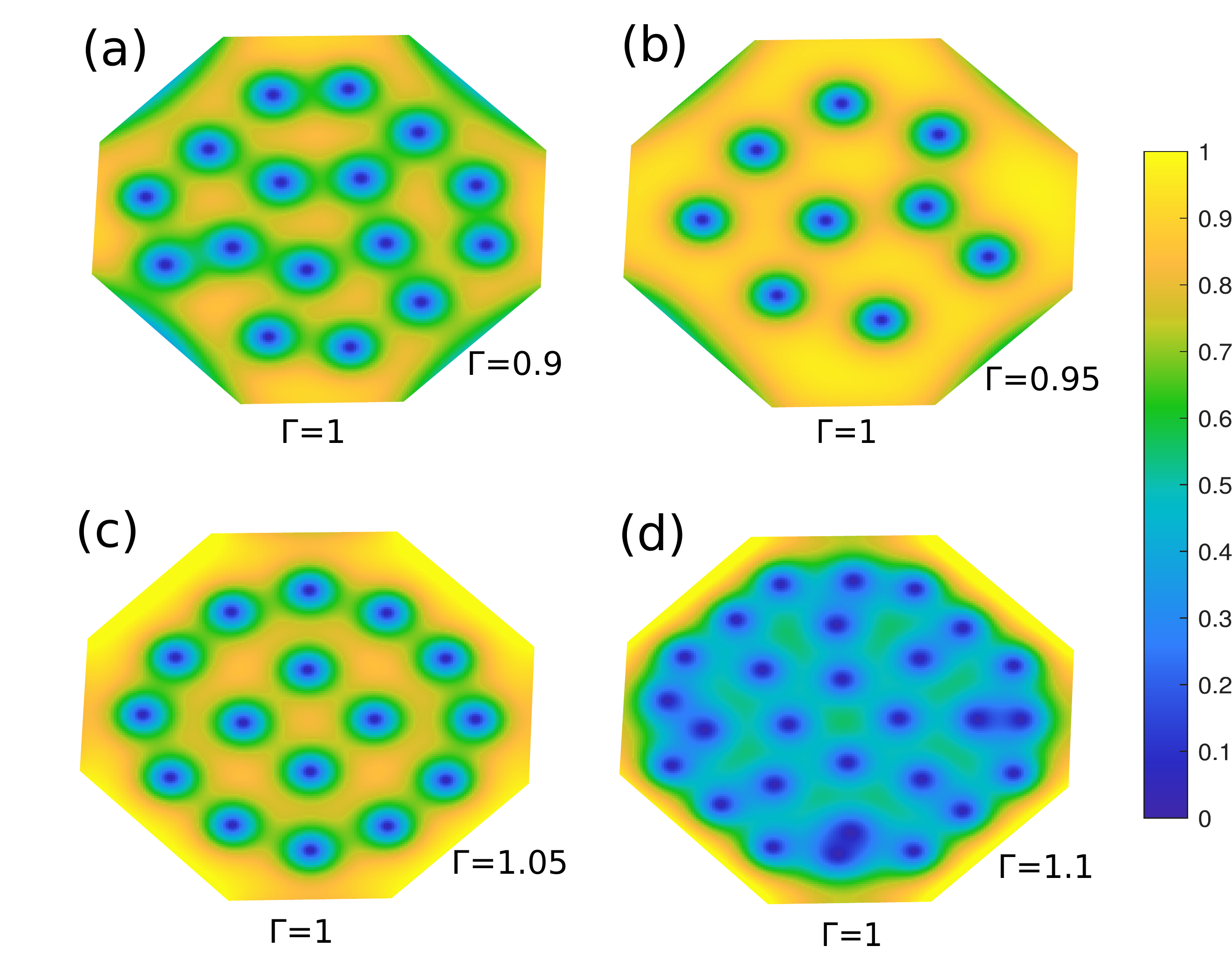}
    \caption{Square order parameter $|\psi|^{2}$ for several boundary conditions for the first configuration of vortex state  at $H_{1}$. (a) $H_{1}(\Gamma=0.90)=0.7$, (b) $H_{1}(\Gamma=0.95)=0.8$, (c) $H_{1}(\Gamma=0.1.05)=0.9$ and (d) $H_{1}((\Gamma=1.10))=1.0$ in four opposite sides of the octagon.}
    \label{dGennes20x20}
\end{figure}
In  Fig. (\ref{dGennes20x20}), we show the sample with $a=20$, and we observe that the vortex configuration is different from that presented in the Fig. (\ref{psi1band20x20})-Fig. (\ref{psi1band30x30}) and additional, the first critical field $H_{1}$ is affected. In Fig. (\ref{dGennes20x20})(a-b) $H=0.7$, $0.8$ and Fig. (\ref{dGennes20x20})(c-d) $H=0.9$, $1.0$, accounting for a change in the lower  field $H_1$, which is precisely the extrapolation effect of the de Gennes parameter in the superconductor state.
Now in Fig. (\ref{M20x20B}) we shown  the magnetization curve $-4\pi M/H_{c2}$ for $\Gamma=0.9;0.95;1.05;1.10$, with which we observe that as $\Gamma$ increases, the critical fields $H_{1}$ and $H_2$ increases, giving the possibility that for lower values of the external field $H$  there is the entry of the vortices in the sample, behavior coinciding with the Fig. (\ref{dGennes20x20}).
In Fig. (\ref{dGhomogeneo}), we present the first stable vortex configuration for the case in which all its boundaries meet the same value $\Gamma$, we found (a) $H_{1}=0.7$, for $\Gamma=0.90$, (b) $H_{1}=0.8$ for $\Gamma=0.95$ (c) $H_{1}=0.9$,$\Gamma=1.05$ and (d) $H_{1}=1.0$ for $\Gamma_{i}=1.1$. Observe that for  $\Gamma<1.0$, $H_{1}$ is lower than for $\Gamma>1.0$, for which there are higher critical fields. Additionally, observe the rotational symmetry in each of the vortex states, see Fig. (\ref{dGhomogeneo}) (a) $C_{1}$, (b) $C_{2}$, (c) $C_{3}$ and (d) $C_{4}$, where all have dihedral planes.
Finally, in the Fig. (\ref{M20x20-H}), we show the magnetization curve for the single-condensate sample, for all border with the same boundary. We found that as $\Gamma$ decreases, $H_{1}$, decreases.
All the calculations previously presented were for the Field Cooling (FC) process. Now, in the Fig. (\ref{ZFCVortex}) we present the first vortex configuration in a ZFC (Zero-Field-Cooling) process, for $\gamma=0$, $a=20\xi$, $b\rightarrow{\infty}$. We observe that as the fixed value of $H$ increases the energy barrier is lower, with which the entry of the vortices in the superconducting sample is favored, this is due to the magnetic pressure in the sample.
Now, in Fig. (\ref{MZFC}) we present the magnetization curves at indicates magnetic fields, in a ZFC process. Especially important is the fact that as $H$ increases, the critical temperature $T_{c}$, for which the system transitions to normal state, is lower. Additionally, with the change in slope, an ingress of vortices in the superconducting system is indicated.
\subsection{Two-condensate case $\gamma=-0.01$, $a=20\xi$ and $\Gamma=1.0$.}\label{Section2b}
In this section we will study the magnetization and the vortex state for the two-condensate system $(\psi_{1},\psi_{2})$, which interaction between them by means of a coupling Josephson type (see Eq. (\ref{Jops})).
\begin{figure}[H]
    \centering
    \includegraphics[scale=0.32]{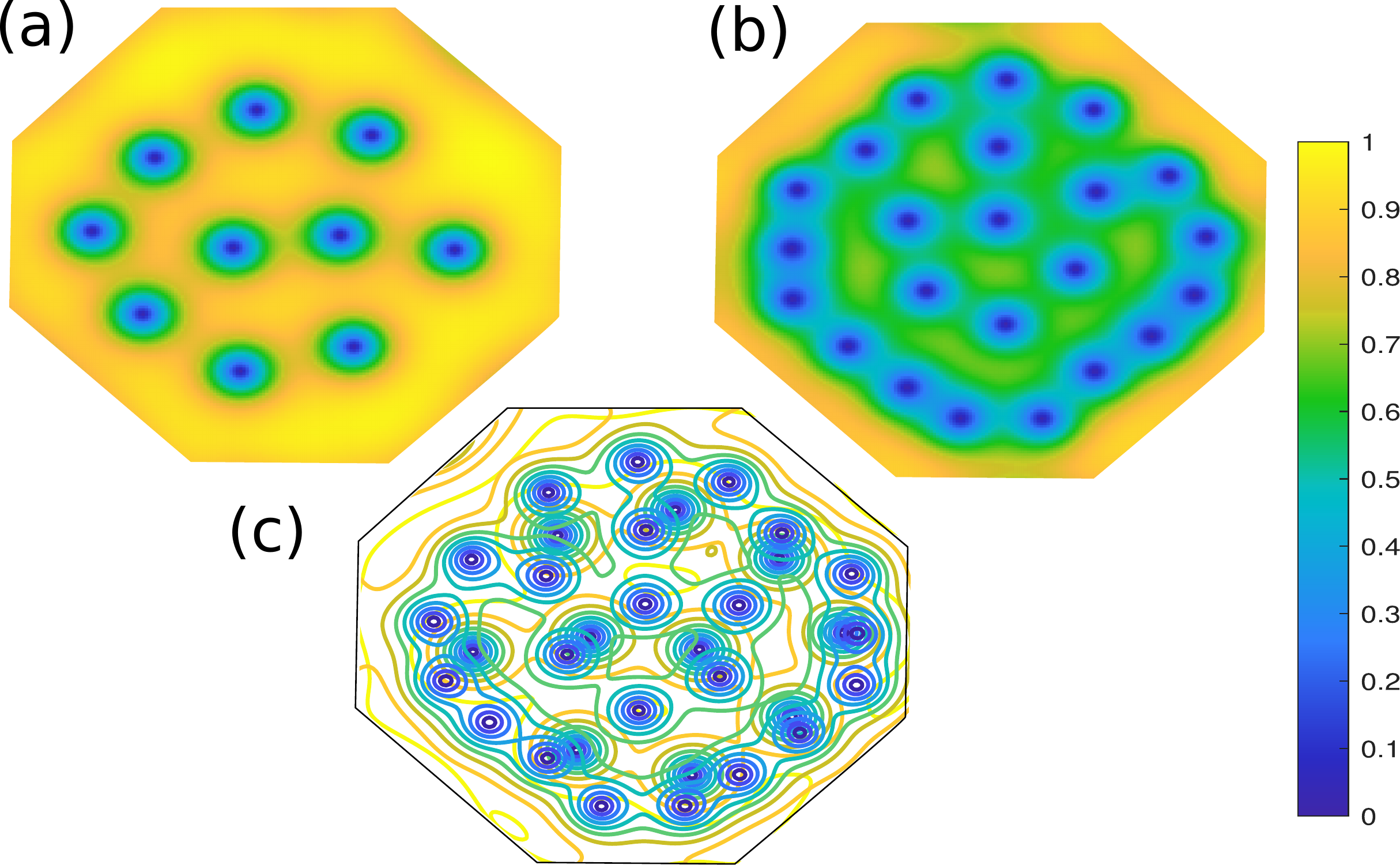}
    \caption{Square order parameter of each condensate (a) $|\psi_{1}|^{2}$, (b)  $|\psi_{2}|^{2}$ and (c) $|\psi_{1}|^{2}$+$|\psi_{2}|^{2}$, for the first stable vortex configuration at $H=H_{1}$. $a=20\xi$ and $b\rightarrow{\infty}$.}
    \label{psi2B}
\end{figure}
In the Fig. (\ref{psi2B}) we present the first stable vortex configuration for (a) the band $1$ $|\psi_{1}|^{2}$, (b) the band $2$ $|\psi_{2}|^{2}$ and (c) the superposition or the bands $|\psi_{1}|^{2}$+$|\psi_{2}|^{2}$. We consider $a=20\xi$ and a superconducting-vacuum interface $b\rightarrow{\infty}$. Thus, we observe that these configurations are different from any of those previously presented. Additionally, it is observed that the vortex number is different in each of them, for (a) $N=10$, (b) $N=23$ and (c) $N=35$. This behavior leads us to think that since all the nuclei of the vortices do not have coincident centers, fractional vortices are formed.
\begin{figure}[H]
    \centering
    \includegraphics[scale=0.51]{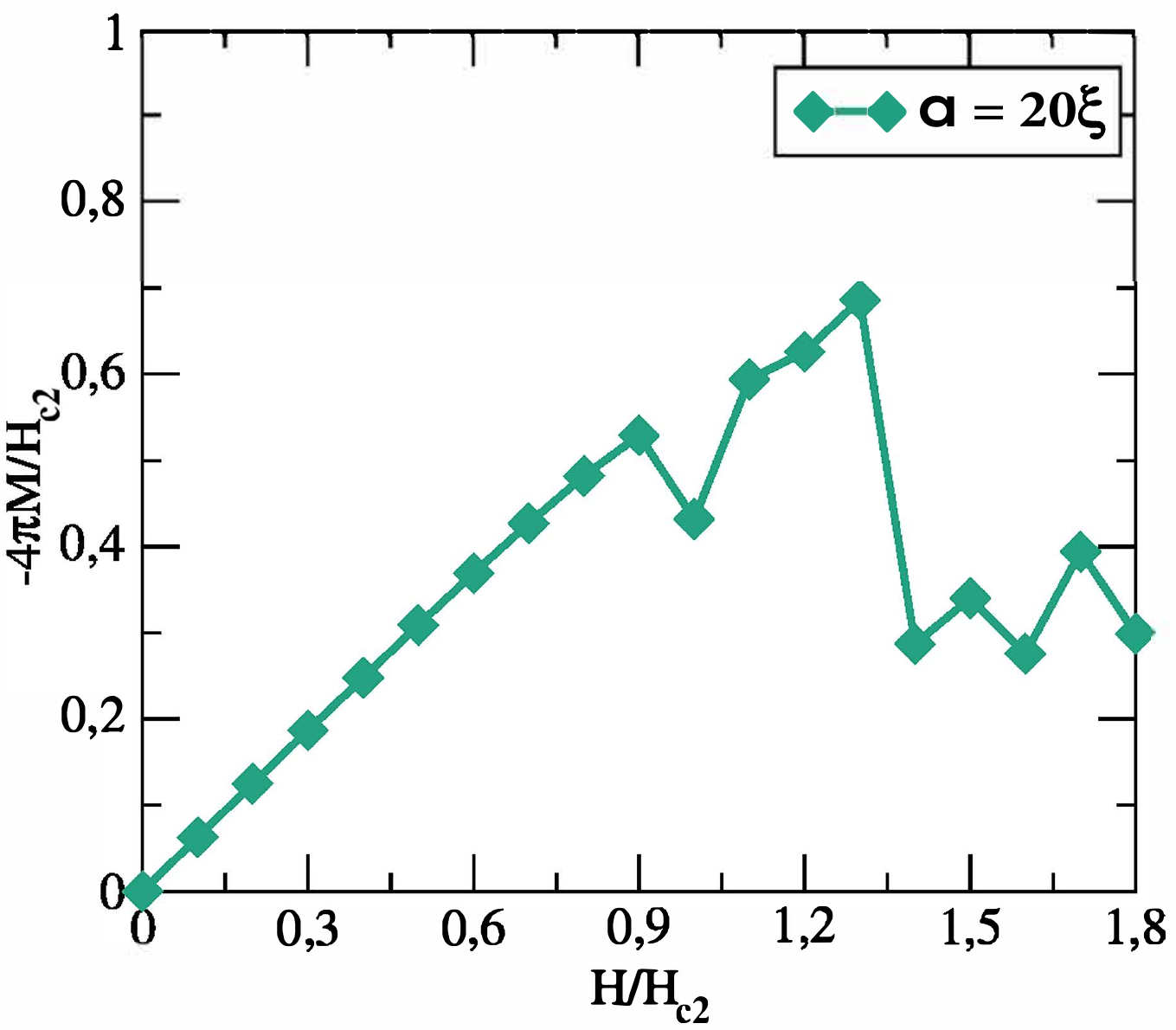}
    \caption{Magnetization $-4\pi M/H_{c2}$ as function of $H$ for the two-condensate sample, with $a=20\xi$ and $b\rightarrow{\infty}$.}
    \label{M20x20}
\end{figure}
Now in the Fig. (\ref{M20x20}), we present the magnetization curve for the two-condensate superconducting system and it is observed that this behavior is non-conventional, since after being in the Meissner-Oschenfeld state $0<H<0.9$, a drop in magnetization is appreciated, where this effect is due to the competition between bands and possible lags ($\phi_{1}-\phi_{2}\not= 0$), which we consider as the first critical point, after that, the Meissner state is retaken for a brief value of the field and for $H=1.2$, it is lost, starting with the mixed state. However, there is an abrupt jump in magnetization, which describes the entry of vortices and then a slight increase again, which describes the behavior of the so-called type $1.5$ superconductor \cite{15Mos}, in which there is a tunneling of vortices between superconducting bands and a non-monotonic interaction between bands (long-range attraction and short-range repulsion), which allows the generation of vortex clusters in the superconducting sample.
\section{Conclusions}\label{Section3}
We have studied a two-dimensional mesoscopic superconducting octagon of one and two condensates that interact with each other, by means of a Josephson-type coupling. The vortex state and the magnetization were studied solving the Ginzburg-Landau equations. We have studied the effects of the de Gennes extrapolation parameter on the critical fields, magnetization and vortex configurations. We have also extended the study of the Abrikosov-Shuvnikov states and magnetization in a two-component sample. Finally, we study the first stable vortex configuration in a Zero Field Cooling  process, we found that the magnetization and critical fields depends strongly on the size of the sample and boundary conditions. Also, we observed that the Abrikosov-Shuvnikov state is not equally distributed in each band, since the number of vortices in each band is different.\\
\section*{ACKNOWLEDGMENTS}\label{Section4}
C. A. Aguirre, would like to thank the Brazilian agency CAPES, for financial support, Grant number: 0.89.229.701-89. J. Faúndez thanks CNPq.


\begin{thebibliography}{}
\bibitem{1} 
\href{https://doi.org/10.1103/PhysRevB.71.024514}
{B.\ J.\ Baelus, K.\ Kadowaki \, and F.\ M.\ Peeters, Phys.\ Rev.\ B, \textbf{71}  024514 (2005)}. 
\bibitem{2} 
\href{https://doi.org/10.1038/s41598-018-36285-4}
{G.\ J.\  Kimmel, A.\ Glatz, V.\ M.\ Vinokur, I.\ A.\  Sadovskyy, Sci.\  Reports, \textbf{9}, 1 (2019).} 
\bibitem{3} \href{https://doi.org/10.1038/373319a0}
{V. V.\ Moshchalkov, L.\ Gielen, C.\ Strunk, R.\ Jonckheere, X.\ Qiu, C.\ Van Haesendonck and Y.\ Bruynseraede, Nature \textbf{373}, 319 (1995).} 
\bibitem{4} 
P.\ G.\ de Gennes, \textit{Superconductivity in Metals and Alloys}, Addison-Wesley, Reading, MA, 1989.
\bibitem{5}
\href{https://doi.org/10.1103/PhysRevLett.12.159}{R. Jaklevic, J. Lambe, A. Silver, and J. Mercereau, Phys.\  Rev.\  Lett.\  \textbf{12}, 159 (1964).}
\bibitem{6}
\href{https://doi.org/10.1103/PhysRevB.72.024529}{Ernst H.\ Brandt,  Phys.\  Rev.\  B \textbf{72}, 024529 (2005).}
\bibitem{7}
\href{https://doi.org/10.1016/j.physc.2007.03.362}{E.\ H.\  Brandt,  Physica C \textbf{327}, 460 (2007).}
\bibitem{8}
\href{https://doi.org/10.1103/PhysRevB.83.174509}{J.\ Carlström, E.\ Babaev, and M.\ Speight, Phys.\  Rev.\  B \textbf{83}, 174509 (2011).}
\bibitem{9}
\href{https://doi.org/10.1103/PhysRevB.85.134514}{M.\ Silaev and E.\ Babaev
Phys.\  Rev.\  B \textbf{85}, 134514 (2012).}
\bibitem{10}
\href{https://doi.org/10.1103/PhysRevLett.105.067003}{E.\ Babaev, J.\ Carlström, and M.\ Speight, Phys.\ Rev.\ Lett. \textbf{105}, 067003 (2010).}
\bibitem{11}
\href{https://doi.org/10.1088/0953-2048/26/7/075005}{F.\  Rogeri, R.\  Zadorosny, P.\  N.\  Lisboa-Filho, E.\  Sardella, and  W.\   Ortiz,  Sup.\   Sci.\  and  Tech.\  \textbf{26}, 075005 (2013).}
\bibitem{12}
\href{https://doi.org/10.1007/s10909-021-02599-3}{C.\   Aguirre,  J.\   Faundez,  S.\   Magalhaes,  A.\   Mosquera-Polo,  and  J.\   Barba-Ortega,  J.\  Low  Temp.\  Phys.\  \textbf{204}, 95 (2021).}
\bibitem{13}
\href{https://doi.org/10.1016/j.physb.2021.413032}{C.\  Aguirre,  A.\   de  Arruda,  J.\  Faundez,  and  J.\   Barba-Ortega,   Physica  B \textbf{615}, 413032 (2021).}
\bibitem{13a}
\href{https://doi.org/10.1103/PhysRevLett.125.137001}{J.\  Tindall, F.\  Schlawin, M.\  Buzzi, D.\  Nicoletti, J.\  R.\  Coulthard, H.\  Gao, A.\  Cavalleri, M.\  A.\  Sentef, and D.\  Jaksch,   Phys.\  Rev.\  Lett.\  \textbf{125}, 137001 (2020).}
\bibitem{14}
\href{https://doi.org/https://doi.org/10.1142/S0217984914502303}{C.\ Aguirre, Q.\ Martins, J.\   Barba-Ortega, Mod.\  Phys.\   Lett.\   B.\  \textbf{33}, 35, 1950435  (2019).}
\bibitem{15}
\href{https://doi.org/10.1016/j.physc.2018.01.016}{X.\  Zhang, G.\  Zhang, L.\  Ying, W.\  Xiong, H.\  Han, Y.\  Wang, L.\  Rong, X.\  Xie, and Z.\  Wang, Physica C \textbf{548},  1  (2018).}
\bibitem{15a}
\href{https://doi.org/10.1103/PhysRevB.100.100503}{E.\  Erlandsen, A.\  Kamra, A.\  Brataas, and A.\  Sudbø, Phys.\  Rev.\  B \textbf{100},  100503(R) (2019).}
\bibitem{15b}
\href{https://doi.org/10.1103/PhysRevB.100.100503}{J.\  Jäykkä, Phys.\  Rev.\  D.\  \textbf{79}, 065006 (2009).}
\bibitem{15c}
\href{https://doi.org/10.1103/PhysRevB.100.100503}{H.\  Nobukane, K.\  Yanagihara, Y.\  Kunisada, Y.\  Ogasawara, K.\  Isono, K.\  Nomura, K.\  Tanahashi, T.\  Nomura, T.\  Akiyama and S.\  Tanda , Sci.\  Rep.\   \textbf{10}, 3462 (2020).}
\bibitem{15d}
\href{https://doi.org/10.1038/s41586-019-1695-0}{X.\ Lu, P.\  Stepanov, W.\ Yang,  Nature \textbf{574}, 653 (2019).}
\bibitem{15e}
\href{https://doi.org/10.1038/s41586-019-1496-5}{D.\  Li, K.\  Lee, B.\  Y.\  Wang, M.\  Osada, S.\  Crossley, H.\  R.\  Lee, Y.\  Cui, Y.\  Hikita, and H.\  Y.\  Hwang, Nature \textbf{572}, 624 (2019).}
\bibitem{15f}
\href{https://doi.org/10.1063/5.0005082}{
P.\ Giannozzi, et.\ al.\, J.\  Phys.\ Condens.\  Matter.\  \textbf{29}, 465901 (2017)}
\bibitem{15g}
\href{https://doi.org/10.1016/j.cpc.2016.07.028}{S.\ Poncé, E.\ R.\  Margine, C.\ Verdi, F.\ Giustino, Computer.\ Phys.\ Communication.\ \textbf{209}, 116 (2016)}
\bibitem{15h} 
\href{https://doi.org/10.1063/1.4819247}
{A.\ C.\ Romaguera and S.\ Silva, J.\ Math.\ Phys\  \textbf{54} 093501 (2013).} 
\bibitem{15i} 
\href{https://doi.org/10.1103/PhysRevLett.107.197001}{J.\ Garaud, J. Carlstr\"om and  E.\ Babaev, Phys.\ Rev.\ Lett.\  \textbf{107} 197001 (2011).} 
\bibitem{15j} 
\href{https://10.1088/1742-6596/490/1/012219}{B.\ Deloof, V.\ V.\ Moshchalkov, L.\ F.\ Chibotaru, J.\ Physics Conf.\ Series, \textbf{490} 012219 (2014).} 
\bibitem{15k}
\href{https://doi.org/10.1103/PhysRevB.72.180502}{ E.\ Babaev and M.\ Speight, Phys.\ Rev.\ B \textbf{72}, 180502(R) (2005).}
\bibitem{15l}
\href{https://doi.org/10.1016/j.physb.2021.413032}{C.\ Aguirre, A.\ S.\ de Arruda, J.\ Faúndez, J.\ Barba-Ortega, Physica B \textbf{615}, 413032 (2021).}
\bibitem{15m}
\href{https://doi.org/10.1103/PhysRevB.70.144523}{L.\ R.\ E.\ Cabral, B.\ J.\ Baelus, F.\ M.\  Peeters, Phys.\ Rev.\ B, \textbf{70} 144523 (2004).}
\bibitem{15n} 
\href{https://doi.org/10.1103/PhysRevB.63.134526}{C.\ C.\ de Souza Silva, L.\ R.\ E.\ Cabral, J.\ Albino Aguiar, Phys. Rev. B \textbf{63}, 134526 (2001)}.
\bibitem{15o}
\href{https://doi.org/10.1103/PhysRevB.83.174509}{ J.\ Carlström, E.\ Babaev, and M.\ Speight, Phys.\ Rev.\ B \textbf{83}, 174509 (2011).}
\bibitem{16}
\href{https://doi.org/10.1016/j.physleta.2015.10.013}{J.\ Barba-Ortega, E.\ Sardella, and J.\ A.\ Aguiar, Phys.\ Lett.\ A.\ \textbf{379}, 732 (2015).}
\bibitem{17} 
\href{https://doi.org/10.1103/PhysRevLett.90.147003}{V.\ R.\ Misko, V.\ M.\ Fomin, J.\ T.\ Devreese, V.\ V.\ Moshchalkov, Phys.\ Rev.\ Lett.\ \textbf{90},147003 (2003).}
\bibitem{18} 
\href{https://doi.org/10.1016/j.ssc.2019.113799}
{C.\ Aguirre, E.\ Sardella, J.\ Barba-Ortega, Solid.\ State.\ Comm.\  \textbf{306}, 113799 (2020).} 
\bibitem{19} \href{https://doi.org/10.1140/epjb/e2020-100418-4}{T.\ Nunes, C.\ Aguirre, A.\ de Arruda and J.\ Barba, Eur.\ Phys.\ J.\ B \textbf{93},  69 (2020).} 
\bibitem{15Mos} \href{https://doi.org/10.1103/PhysRevLett.102.117001}{V.\  V.\ Moshchalkov, M.\ Menghini, T.\  Nishio, Q.\  H.\  Chen, A.\  V.\  Silhanek, V.\  H.\  Dao, L.\  F.\  Chibotaru, N.\  D.\  Zhigadlo, and J.\  Karpinski, Phys.\ Rev.\ Lett.\ 102, 117001 (2009)}
\end{thebibliography}
\end{document}